\documentclass[fleqn,usenatbib]{mnras}

\usepackage{newtxtext,newtxmath}

\usepackage[T1]{fontenc}

\DeclareRobustCommand{\VAN}[3]{#2}
\let\VANthebibliography\thebibliography
\def\thebibliography{\DeclareRobustCommand{\VAN}[3]{##3}\VANthebibliography}
\usepackage{scalerel}
\usepackage{tikz}
\usetikzlibrary{svg.path}

\definecolor{orcidlogocol}{HTML}{A6CE39}
\tikzset{
  orcidlogo/.pic={
    \fill[orcidlogocol] svg{M256,128c0,70.7-57.3,128-128,128C57.3,256,0,198.7,0,128C0,57.3,57.3,0,128,0C198.7,0,256,57.3,256,128z};
    \fill[white] svg{M86.3,186.2H70.9V79.1h15.4v48.4V186.2z}
                 svg{M108.9,79.1h41.6c39.6,0,57,28.3,57,53.6c0,27.5-21.5,53.6-56.8,53.6h-41.8V79.1z M124.3,172.4h24.5c34.9,0,42.9-26.5,42.9-39.7c0-21.5-13.7-39.7-43.7-39.7h-23.7V172.4z}
                 svg{M88.7,56.8c0,5.5-4.5,10.1-10.1,10.1c-5.6,0-10.1-4.6-10.1-10.1c0-5.6,4.5-10.1,10.1-10.1C84.2,46.7,88.7,51.3,88.7,56.8z};
  }
}

\newcommand\orcidicon[1]{\href{https://orcid.org/#1}{\mbox{\scalerel*{
\begin{tikzpicture}[yscale=-1,transform shape]
\pic{orcidlogo};
\end{tikzpicture}
}{|}}}}

\usepackage{graphicx}
\usepackage{txfonts}

\usepackage[export]{adjustbox}
\usepackage{amsmath}	
\usepackage{amssymb}	
\usepackage{subfig}
\usepackage{fancyhdr}
\usepackage{url}
\usepackage{float}
\usepackage{multirow}
\usepackage{appendix}
\usepackage{framed}
\usepackage{tcolorbox}
\usepackage{adjustbox}
\usepackage{stackengine}
\usepackage{framed}
\usepackage{multicol}
\usepackage{hyperref}
\usepackage{siunitx}

\usepackage{graphicx}	
\usepackage{amsmath}	

\title[(130) Elektra Delta - on the stability of the new third moonlet]{(130) Elektra Delta - on the stability of the new third moonlet}

\author[G. Valvano; R. Machado Oliveira; O. C. Winter;  R. Sfair; G. Borderes-Motta;]
        {G. Valvano$^{1}$\thanks{E-mail:  giulia.valvano@unesp.br}\orcidicon{0000-0002-7905-1788}\,
        R. Machado Oliveira$^{1}$\thanks{E-mail: rai.machado@unesp.br }\orcidicon{0000-0002-6875-0508}\,
        O. C. Winter$^{1}$\thanks{E-mail:  othon.winter@unesp.br}\orcidicon{0000-0002-4901-3289}\,
        R. Sfair$^{1,}$ $^{2}$\thanks{E-mail: rafael.sfair@unesp.br}\orcidicon{0000-0002-4939-013X}
    \newauthor
        G. Borderes-Motta$^{3,}$ $^{4}$\thanks{E-mail: gabriel.borderes@uc3m.es}\orcidicon{0000-0002-4680-8414}\
\\
$^{1}$ Grupo de Din\^amica Orbital e Planetologia, S\~ao Paulo State University, UNESP, Guaratinguet\'{a}, CEP 12516-410, 
  S\~{a}o Paulo, Brazil\\
$^{2}$ Institut für Astronomie und Astrophysik, Eberhard Karls Universität Tübingen, Germany\\
$^{3}$ Bioengineering and Aerospace Engineering Department, Universidad Carlos III de Madrid, Leganés, 28911, Madrid, Spain\\
$^{4}$ Swedish Institute of Space Physics, IRF, Kiruna, 98128, Norrbotten, Sweden}

\date{Accepted XXX. Received YYY; in original form ZZZ}

\pubyear{2015}

\begin{document}
\label{firstpage}
\pagerange{\pageref{firstpage}--\pageref{lastpage}}
\maketitle

\begin{abstract}
The aim of this work is to verify the stability of the proposed orbital solutions for the third moonlet (Delta) taking into account a realistic gravitational potential for the central body of the quadruple system (Alpha). We also aim to estimate the location and size of a stability region inside the orbit of Gamma. First, we created a set of test particles with intervals of semi-major axis, eccentricities, and inclinations that covers the region interior to the orbit of Gamma, including the proposed orbit of Delta and a wide region around it. We considered three different models for the gravitational potential of Alpha: irregular polyhedron, ellipsoidal body and oblate body. For a second scenario, Delta was considered a massive spherical body and Alpha an irregular polyhedron. Beta and Gamma were assumed as spherical massive bodies in both scenarios. The simulations showed that a large region of space is almost fully stable only when Alpha was modeled as simply as an oblate body. For the scenario with Delta as a massive body, the results did not change from those as massless particles. Beta and Gamma do not play any relevant role in the dynamics of particles interior to the orbit of Gamma. Delta's predicted orbital elements are fully unstable and far from the nearest stable region. The primary instability source is Alpha's elongated shape. Therefore, in the determination of the orbital elements of Delta, it must be taken into account the gravitational potential of Alpha assuming, at least, an ellipsoidal shape.

\end{abstract}




\section{Introduction}
Asteroids with companion bodies are particularly remarkable,
since their formation process may provide insight into the evolution 
of our Solar System.
The origin of these multiple systems can be
by collisions, accretion, disruption, or other processes
\citep{vokrouhlicky2008pairs, margot2015asteroid}. 
The knowledge about the formation process
also may contribute to the information on
compositions, bulk density, interior structure, 
dynamics, and mechanics of their
evolutionary processes and of our Solar System. 

The (130) Elektra, a quadruple system, is formed by the primary body,
named here as Alpha, 
and three moonlets: the outer and larger Beta (S/2003 (130) 1), 
Gamma (S/2014 (130) 1), and  the inner and 
smaller satellite Delta (S/2014 (130) 2).
The first satellite Beta 
was discovered  in 2003
\citep{2003IAUC.8183....1M}. With observational 
data obtained  in the same year,
\citet{marchis2006shape} modeled a shape model 
for Elektra with a mean 
diameter of 191 km, and estimated a radius of 3 km
for the secondary body considering the Elektra's low albedo 
as $\sim$0.5. \citet{marchis2008main} 
reported a radius of about 3.5 km for Elektra's 
moonlet S/2003 (130) 1 and the 
orbital eccentricity of $\sim$0.1, probably caused by tidal effects. 
They also estimated the mass of the system 
as $6.6 \pm0.4 \times10^{18}$ kg.
From the adaptive optics campaign, \citet{marchis2008main} 
did not identify any additional  
 bodies in the Elektra system.
However, \citet{yang2016extreme} through 
direct imaging and integral field
spectroscopy detected 
a second moonlet named S/2014 (130) 1 (Gamma). 
It is estimated to have about 1 km in 
radius and an eccentric orbit, 
but due to the elongation of Elektra, the orbital solutions 
were not strictly constrained.

A careful analysis and treatment of the data led to 
the discovery of the moon S/2014 (130) 2 (Delta), making 
(130) Elektra the first known quadruple asteroid system.
\citet{refId0} applied new techniques of adaptive optics
to obtain astrometric data of the system. They modeled 
the asteroids' halo, extracted the moons' signals and 
their positions at different epochs, deriving the orbital
elements of the moonlets and solutions of the orientation 
of the poles. Due to the 
inherent ambiguity of the data, \citet{refId0} found some
numerical solutions that include polar and retrograde
orbits for the satellites. However, the authors argued
that the fit resulting in prograde motions of the three
satellites is physically more plausible.

\citet{refId0} pointed out that some uncertainties still
remain in the orbit of Delta. 
Considering the best orbital fit reported, we aim to verify the
orbital stability of the new moon through numerical simulations
taking into account the irregular shape of the main body.
A significant instability is verified in the region where the new moonlet
is supposed to be. We initially present the system and 
the central body's irregular shape (section \ref{system}). 
In section \ref{stability}, we present a map of 
stability for the orbital solution of the third moonlet 
as a massless particle. Moreover, 
we discussed the stability of the third moon as
a massive body. Our final comments are
in section \ref{final}.

\section{The quadruple asteroid system}
\label{system}

\begin{figure}
\begin{center}
\includegraphics*[trim = 0mm 3cm -0.1cm 0mm, width=.9\columnwidth]{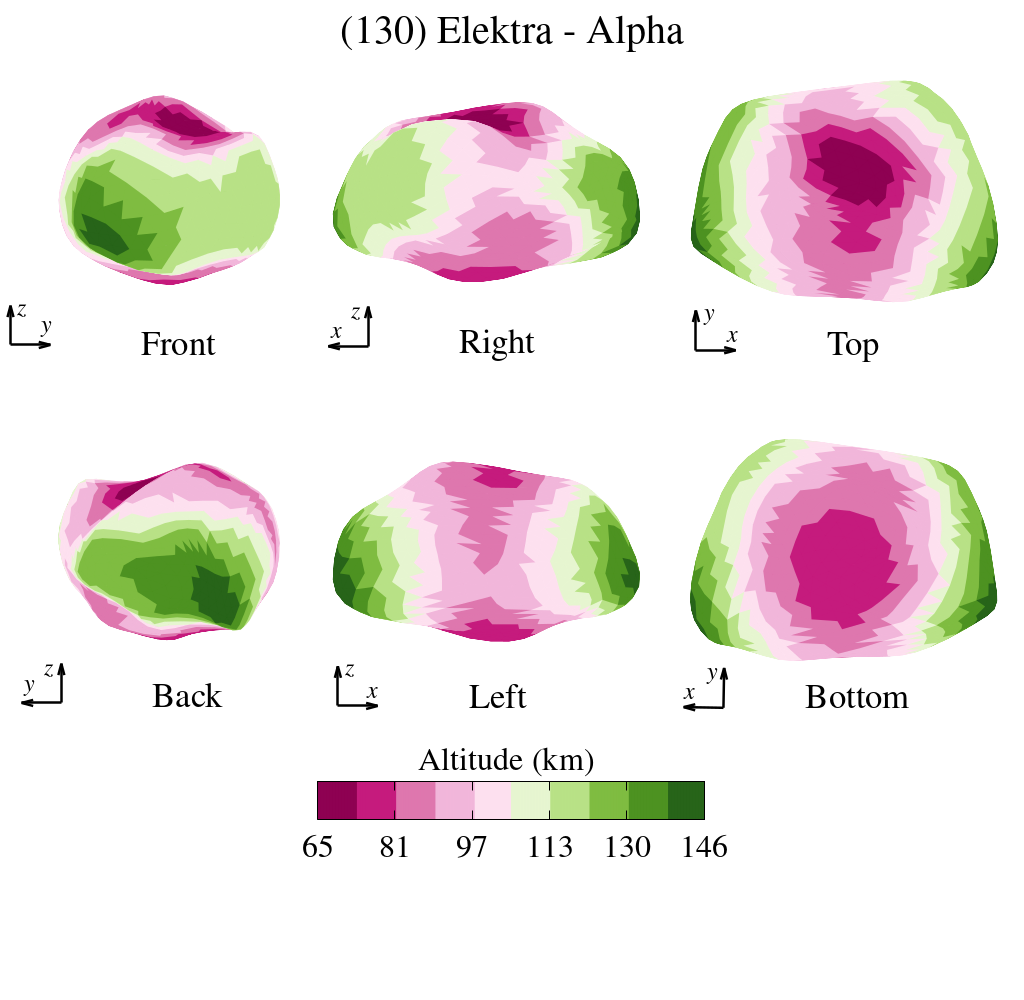}
\end{center}
\caption{Altitude mapped across the surface of Alpha, the primary body of the (130) Elektra,  under different views.}
\label{fig:geom}
\end{figure}

The main body of the system, Alpha, has an elongated and flattened shape 
as indicated by its
gravitational coefficients whose values are 
J$_2$= 0.1852 and C$_{22}$= 0.0448 (normalized by
the spherical radius of 98~km).
Alpha has a spin period of 5.224663 ~h \citep{hanuvs2017shape}, 
resulting in a radius of corotation of $\sim 158$ km.

From the polyhedral model with 1800 triangular
faces and 902 vertices \citep{hanuvs2017shape} 
we mapped the altitude of Alpha (Fig. \ref{fig:geom}). 
The distribution of altitude shows that the 
minimum altitude of Alpha is located in
the poles, while the maximum values are in the 
equator. One can note a depression 
in the north pole, while the
south there is a mountainous region.

\begin{figure}
\begin{center}
\includegraphics*[trim = 0mm 0cm -0.1cm 0mm, width=.9\columnwidth]{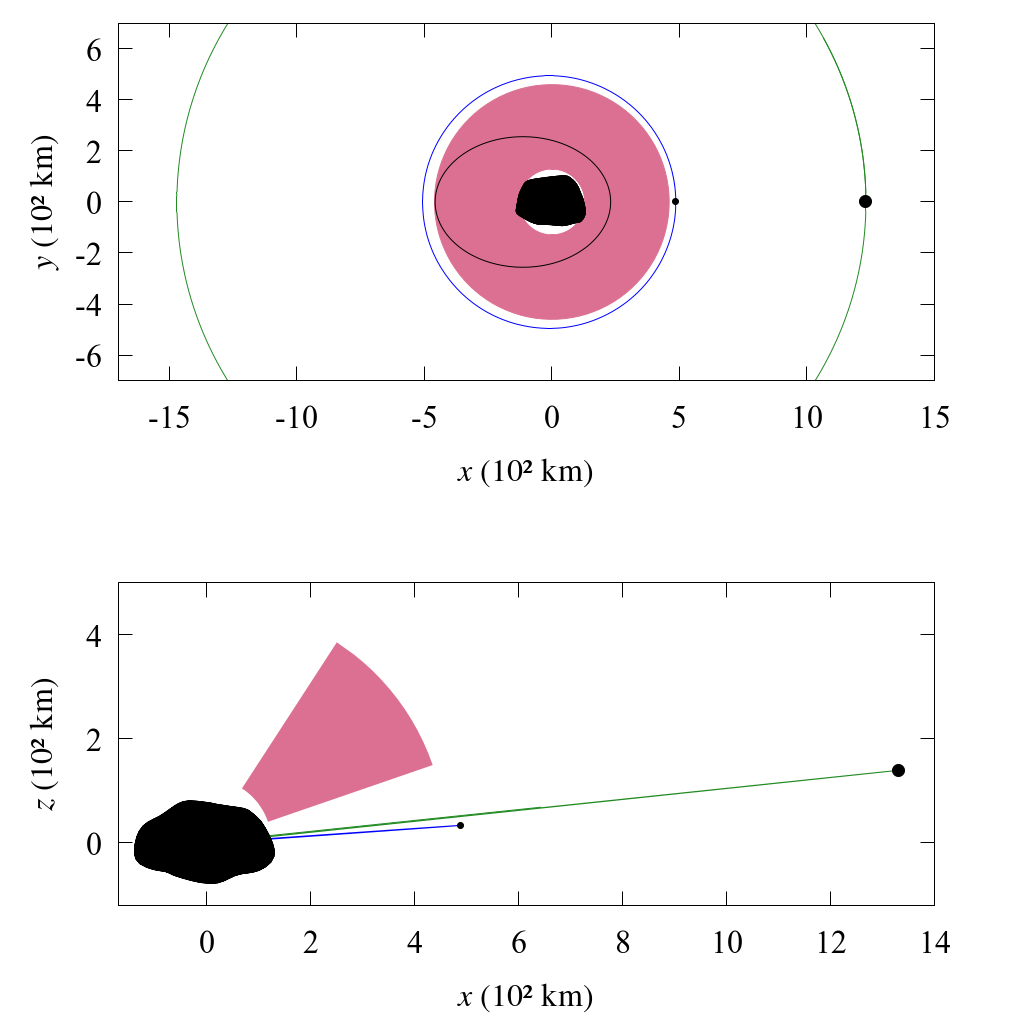}
\end{center}
\caption{Representation of the (130) Elektra system in the Alpha equatorial plane $xoy$ (upper panel) 
and the $xoz$ plane to show an inclination perspective (lower panel). 
The bodies are black with Alpha's polyhedral format in scale. 
The pink region corresponds to the initial condition of the particles. 
The orbits of Beta and Gamma are shown in green and blue lines, respectively, and  
the black line represents the nominal orbit of Delta.
The upper panel shows the Top view of Alpha's format, and the lower
panel, the Left view (see Fig. \ref{fig:geom} for more details).}
\label{fig:orbit}
\end{figure}

Figure \ref{fig:orbit} shows the system from top and side views, 
where the $xoy$ plane corresponds to Alpha's equatorial plane,
and an inclination projection considering the $xoz$ plane. 
The green, blue and black lines represent the orbits of 
Beta, Gamma, and Delta, respectively. The orbits of the moonlets
in Fig. \ref{fig:orbit} correspond to the nominal values 
(see Table~\ref{table: infos}). 
Alpha is represented by its irregular shape
model in scale, while Beta and Gamma are shown as black dots. 
Delta has an eccentricity about one
order of magnitude larger than the other two moonlets and 
has the largest inclination (Table \ref{table: infos}). 
However, these orbital values are uncertain due to
the observation data \citep{refId0}. Despite that, it 
is noticeable that the  projected apocentre of the orbit of Delta is  
close to Gamma,
that could produce  an instability. 
The other physical and orbital data that we 
adopted in this work are shown in Table \ref{table: infos}.

\begin{table*}[!h]
\centering
\caption{Orbital and physical data of each component of the system.}
\label{table: infos} 
\begin{tabular}{c|c|c|c|c|c|c|c|c}
\hline\hline
Body & Volumetric & Density$^{b}$ & Mass & Orbit & Semi-major & Eccentricity & Inclination & Orbital\\
  & diameter (km) & (g$\cdot$cm$^{-3}$) & (10$^{18}$ kg) & & axis &  & (deg)$^{e}$ & period\\
\hline
\hline
Alpha & 199.0$^{b}$ & 1.6 & 6.60$^{c}$ & Sun & 3.13 au$^{d}$ & 0.21$^{d}$  & 22.78$^{d}$  & 5.53 years$^{d}$\\
Beta & 6.0$^{a}$ & 1.6 & 1.81 $\times$ 10$^{-4}$& Alpha & 1353.0 $\pm$ 17 km$^{a}$ & 0.09$^{a}$ & 6 $\pm$ 1$^{a}$ & 5.287 days$^{a}$\\
Gamma & 2.0$^{a}$ & 1.6 &  6.70 $\times$ 10$^{-6}$& Alpha & 501.0 $\pm$ 7 km$^{a}$ & 0.03$^{a}$ & 4 $\pm$ 5$^{a}$ & 1.192 days$^{a}$\\ 
Delta & 1.6$^{a}$ & 1.6 & 3.43 $\times$ 10$^{-6}$& Alpha & 344.0 $\pm$ 5 km$^{a}$ & 0.33 $\pm$ 0.05$^{a}$ & 38 $\pm$ 19$^{a}$ & 0.679 days$^{a}$\\ 
\hline\hline
\end{tabular}
  \begin{flushleft}
  	\quad  {\footnotesize $^{a}$ \citet{refId0}}
  				
  	\quad  {\footnotesize $^{b}$ \citet{hanuvs2017shape}}
  	
  	\quad  {\footnotesize $^{c}$ \citet{marchis2008main}}
  				
  	\quad  {\footnotesize $^{d}$ Johnstons Archive. Website: \href{http://www.johnstonsarchive.net/astro/astmoons/am-00130.html}{http://www.johnstonsarchive.net}}
  	
  	  \quad  {\footnotesize $^{e}$ The inclination of Alpha's orbit is with respect to the ecliptic plane. The inclination of the moonlets is with respect to the Alpha's equator.}
\end{flushleft}
\end{table*}

Since the mass of each satellite is not known and no noticeable 
spectral difference was perceived among the system objects 
\citep{yang2016extreme}, we assumed that they have the same 
bulk density as Alpha and, as a consequence, the masses  
considering spherical bodies with the diameters
reported by \citet{hanuvs2017shape} and \citet{refId0} 
(Table \ref{table: infos}). The mass of Beta is about four
orders of magnitude  smaller than Alpha's and two 
orders of magnitude larger than Gamma's, and 
Beta is about three times  further from Alpha than Gamma. Thus,
its gravitational perturbation may be too small in
the orbit of Delta. Gamma and Delta are closer to each
other, but  Alpha's mass is six orders of magnitude larger.
Therefore, the orbit of Delta is dominated by the gravitational contribution of 
Alpha. 
This analysis will be confirmed in section \ref{stability}, 
where it is shown that the contribution of Beta and Gamma are almost 
irrelevant for the orbital evolution of bodies in trajectories interior 
to that of Gamma.

\section{Stability}
\label{stability}

We computed the Roche radius \citep{murray1999solar} 
using
the parameters presented in Table \ref{table: infos} and 
obtained a radius of 143.5 km. Thus, Delta could exist
in the region where it is reported to be
without being disintegrated by the
tidal forces from Alpha.

To explore the stability of Delta, 
we performed numerical simulations with the N-BoM package \citep{winter2020asteroid}.
In our first set of simulations, we adopted the mascons method 
\citep{Geissler1996,borderes-motta18} to compute 
the gravitational potential of Alpha.
The mass of Alpha was uniformly 
divided into 33,035 mascons equally spaced into a
tridimensional grid following its polyhedral shape model. 
The  moonlets Beta and Gamma
were considered as spherical massive bodies.
We investigated two scenarios: 
one assuming  Delta as a massless particle, and  other considering its mass.
The following sub-sections present the stability analysis for each case.

\subsection{Massless particle}
\label{massless}

\begin{figure*}
\begin{center}
\subfloat[]{\includegraphics*[trim = 0mm 0cm 0cm 0mm,
width=1.95\columnwidth]{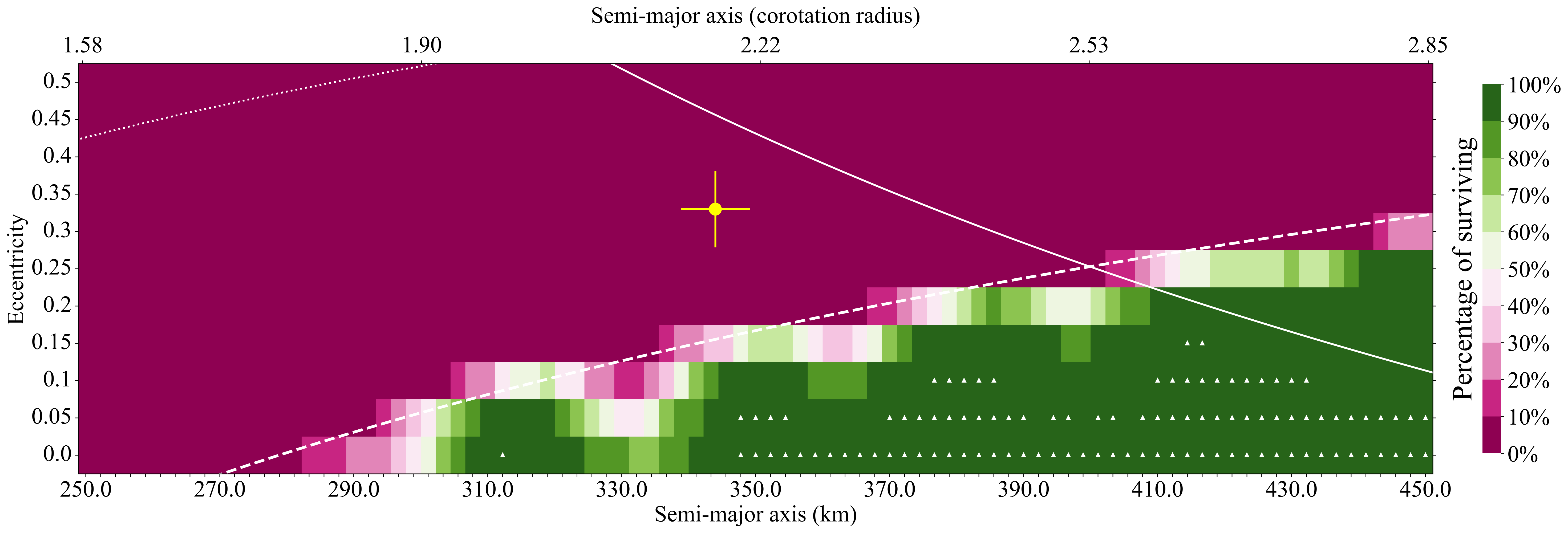}\label{afig:MapaSemMassa}}\\
\subfloat[]{\includegraphics*[trim = 0mm 0cm 0cm 0mm,
width=1.95\columnwidth]{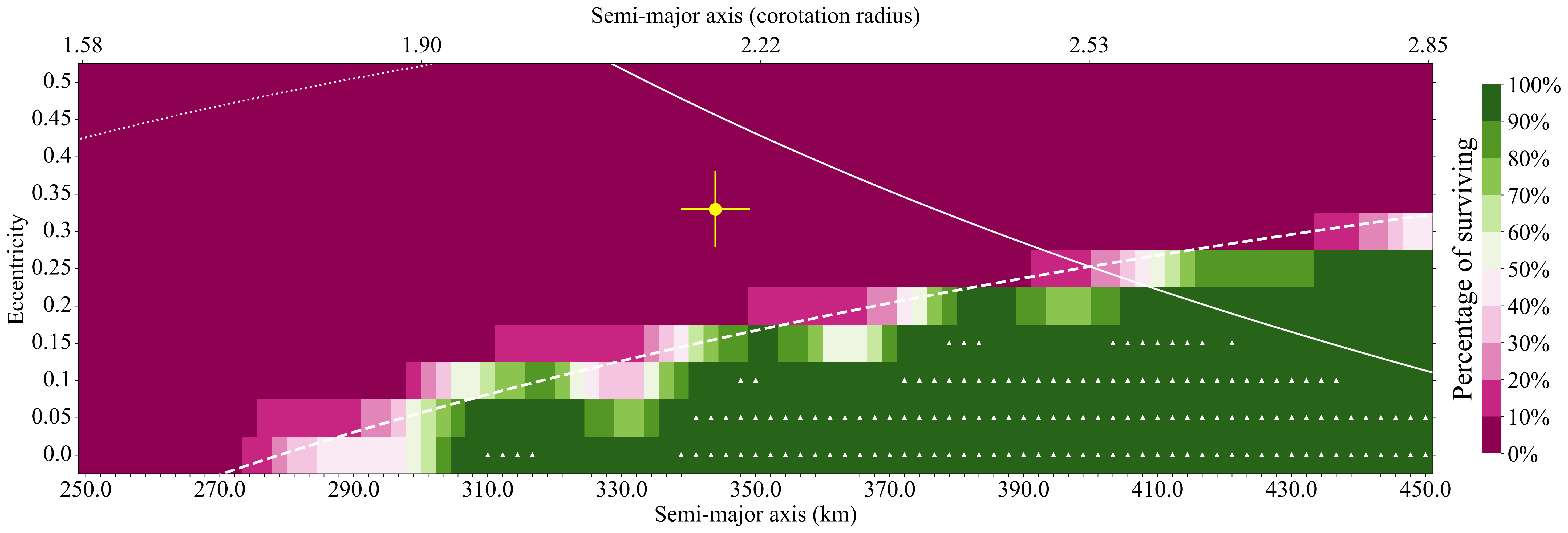}\label{bfig:MapaSemMassa}}\\
\subfloat[]{\includegraphics*[trim = 0mm 0cm 0cm 0mm,
width=1.95\columnwidth]{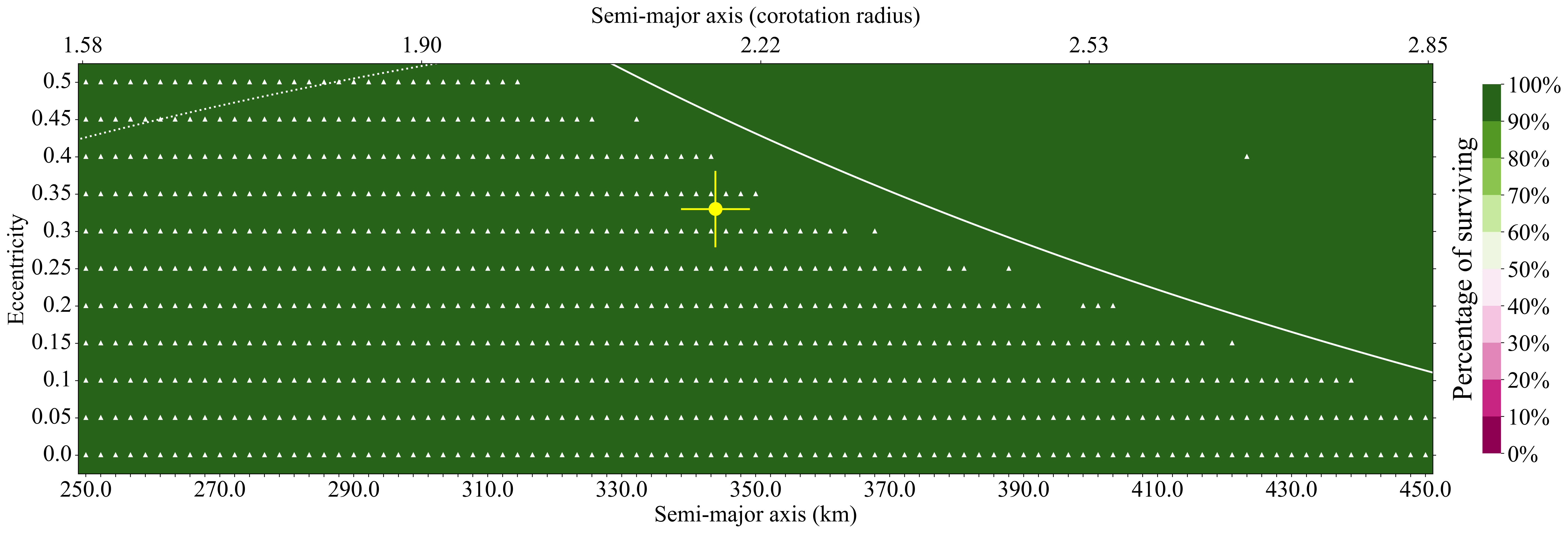}\label{cfig:MapaSemMassa}}
\end{center}
\caption{\label{fig:MapaSemMassa} Stability diagram for five years of integration in the region interior to the orbit of Gamma. The color bar represents the percentage of surviving particles. The white triangles indicate 100\% of survival. The yellow cross corresponds to orbital elements of Delta with the respective error bars, 
while the yellow dot represents the nominal elements of Delta.
The full white line indicates the collision line with Gamma. The dotted white line indicates the Roche limit pericenter curve.
The dashed white line represents the chaotic limiting border given by the analytical expression of \citet{lages2017chaotic, lages2018chaotic}.}
Each diagram represents a different gravitational potential model for Alpha: (a) the full irregular shape (using mascons), (b) the ellipsoidal shape (using mascons), and (c) the oblate body shape (including the term with J$_2$).  For all the simulations, Beta and Gamma were considered as spherical massive bodies.
\end{figure*}

In this work, we performed numerical simulations 
to analyse the stability of the  region
interior to the orbit of Gamma.
In that way, we are covering more than the observational uncertainties for
the orbital elements of Delta
as shown in Table \ref{table: infos} and exemplified
by the pink region of Fig. \ref{fig:orbit}. 
To prevent overlooking small regions that could
be stable, we adopted a fine grid of initial conditions
to explore a robust set of initial conditions. 
For the initial conditions of the test particles, 
we assumed a semi-major axis ($a$) in the range of
[250, 450] km in steps
of 200 m and the eccentricity ($e$)
from 0 to 0.5 in steps of 0.05. 
Each ($a$, $e$) pair has 100 particles with 
inclinations randomly distributed in the interval
 \ang{19} to \ang{57} as proposed by \citet{refId0}. 
  The initial conditions of the mean longitude, the longitude of
ascending node and the longitude of pericentre were randomly 
chosen between 0$^{\circ}$ to 360$^{\circ}$.
 As a result of
 these combinations of eccentricities and 
 semi-major axis, we had a set of 1,101,100
 particles located inside the orbit of Gamma
(the pink region in Fig. \ref{fig:orbit}). 
The system was numerically integrated for five years
($\sim$ 2,700 orbital periods of Delta).
In all simulations, Beta and Gamma were considered  as spherical massive bodies.
We consider the semi-major axis, inclination, and eccentricity presented
in Table \ref{table: infos} to create the initial conditions of Beta and Gamma.  
We set their initial values of mean longitude, longitude of ascending node
and longitude of pericentre as  0$^{\circ}$. For this reason, Beta and Gamma
were initially aligned over the line of nodes.
Table \ref{table: xyz} gives the state vectors of Beta and
Gamma that were used in all simulations.
The results are presented in terms of 
 an ($a$, $e$) stability diagram 
colored according
to the percentage of surviving  particles (Fig. \ref{fig:MapaSemMassa}).

    \begin{table*}
    \centering
    \caption{The initial state vectors of Beta and Gamma relative to Alpha (origin of the system).}
    \label{table: xyz} 
    \begin{tabular}{c|c|c|c|c|c|c}
    \hline\hline
    Body & $x$ (km) & $y$ (km) & $z$ (km) & $v_x$ (m$\cdot$s$^{-1}$) & $v_y$ (m$\cdot$s$^{-1}$) & $v_z$ (m$\cdot$s$^{-1}$)\\
    \hline
    \hline
    Beta & 1.23123e+03 & 0.0 & 0.0 & 0.0 & 1.96365e+01 & 2.06388\\
    Gamma & 4.85970e+02 & 0.0 & 0.0 & 0.0 & 3.04757e+01 & 2.13107\\
    \hline\hline
    \end{tabular}
    \end{table*}

According to the results shown in Fig. \ref{fig:MapaSemMassa}a,
there is  a large 
instability region, and  particles with
semi-major axes smaller than 300 km are mostly removed. 
The same occurs to  the particles with 
eccentricities larger than $0.3$, regardless of the initial semi-major axis. 
From the unstable particles, 80\% of them exceeded
the ejection distance of 13,530.0 km ($\sim10$ times
the value of the nominal semi-major axis of Beta),
while 20\% collided with one of the bodies. 
The vast majority of the collisions occurred with
Alpha, while only 0.72\% of these collisions occurred with
Gamma and no particles collided with Beta. 
Among all the removed particles, 80\% of them were ejected or collided in the first six months of integration.
In Fig. \ref{fig:MapaSemMassa} the full white line
denotes the collision line with Gamma, indicating
the limit of which the particle reaches the projected orbit of Gamma. 
It is important to note that the inclination between the orbital planes of Gamma and Delta is very high. 
Thus,  this justifies why only a few collisions were recorded, 
pointing out that the 
shape of Alpha is the main source of instability.
Nevertheless, there are stable regions with a
semi-major axis larger than
300 km and eccentricities smaller than 0.25. 
For particles in these regions, the pericentre do
not have close approaches with Alpha, and their orbits do
not become unstable for 5 years. 

The yellow dot in Fig.~\ref{fig:MapaSemMassa}  represents the nominal elements of Delta 
with the corresponding uncertainties.  
Thus, when treated as a massless object, Delta is not
located in a stable region with the reported orbital elements. 
Assuming that the semi-major axis of 344 km is
correct, the eccentricity should be at least a factor of 3 smaller
to be in a stable region.  Conversely, there is no stable semi-major axis 
for the nominal eccentricity. 
Therefore, Delta's proposed orbital elements are far from any stable region.

In order to get some hint on the relevance of the shape of Alpha
to the stability of Delta,
 we considered two simplified models for Alpha's  gravitational potential. 
One assuming a flattened and elongated ellipsoidal body shape based on the equivalent dimensions  of the irregular polyhedron shape model (semi-axes: $a$=137.5 ~km, $b$=100 ~km, $c$=73.3 ~km).
This gravitational potential was simulated using about 10,557 mascons, 
obtained similarly to the fully irregular shape case.
The other model was even simpler. Alpha was assumed to be an oblate body.
It was simulated using the $J_2$ term included in its  gravitational potential.
 In both approaches, Beta and Gamma were considered massive spherical bodies.
We performed the whole set of simulations for these two models of Alpha.
Figure \ref{fig:MapaSemMassa}b presents the diagram of stability for the ellipsoidal model.
The overall picture is similar to the results found for the 
model of the irregular shape of Alpha
(Figure \ref{fig:MapaSemMassa}a), with only subtle differences.

\citet{yang2016extreme} found admissible orbital 
solutions for a range of $J_2$ values between 0 and 0.13. 
However, 
they concluded that more observational data was necessary to 
constrain  the coefficient. Here we assume  
$J_2$=$0.1852$ calculated according to
\citet{huscheeres2004} by using the 
principal moments of inertia computed
using the algorithm of \citet{mir1996}.
Our results, considering the oblate format of Alpha,
reproduced by the addition of the J$_2$
coeﬃcient term in the gravitational potential, show a
completely different scenario (Figure \ref{fig:MapaSemMassa}c) 
from the  two previous cases (Figure \ref{fig:MapaSemMassa}a, b).
The vast majority of particles in the whole region is stable, 
including the suggested range of trajectories associated with the moonlet Delta.

\citet{lages2017chaotic, lages2018chaotic} made studies 
considering a dumbbell
body of size $d$ modeled by two spheres of masses $m_1$ and $m_2$
around its center of mass with a Keplerian rate of rotation $\omega_0$
and an angular frequency $\omega$. 
They showed that a chaotic limiting border
around a rotating gravitating dumbbell can be analytically estimated.
In order to make a comparison between their approach and our simulations, 
we used their theory considering
Alpha as a dumbbell body given by two
equal masses and a radius of 100 km, with an angular frequency
$\omega$=$0.99907\omega_0$.
The dashed lines in 
Fig. \ref{fig:MapaSemMassa}a and \ref{fig:MapaSemMassa}b show
the border of the chaotic region around Alpha.
The region above the border is the chaotic region. 
Note that the nominal orbital
elements of Delta lie in the
chaotic region. The region below the limiting border is expected to be stable, 
which is in agreement with our results that show 
a high percentage of survivors.

\begin{figure*}
\begin{center}
\includegraphics*[trim = 0mm 0cm 0cm 0mm,
width=1.95\columnwidth]{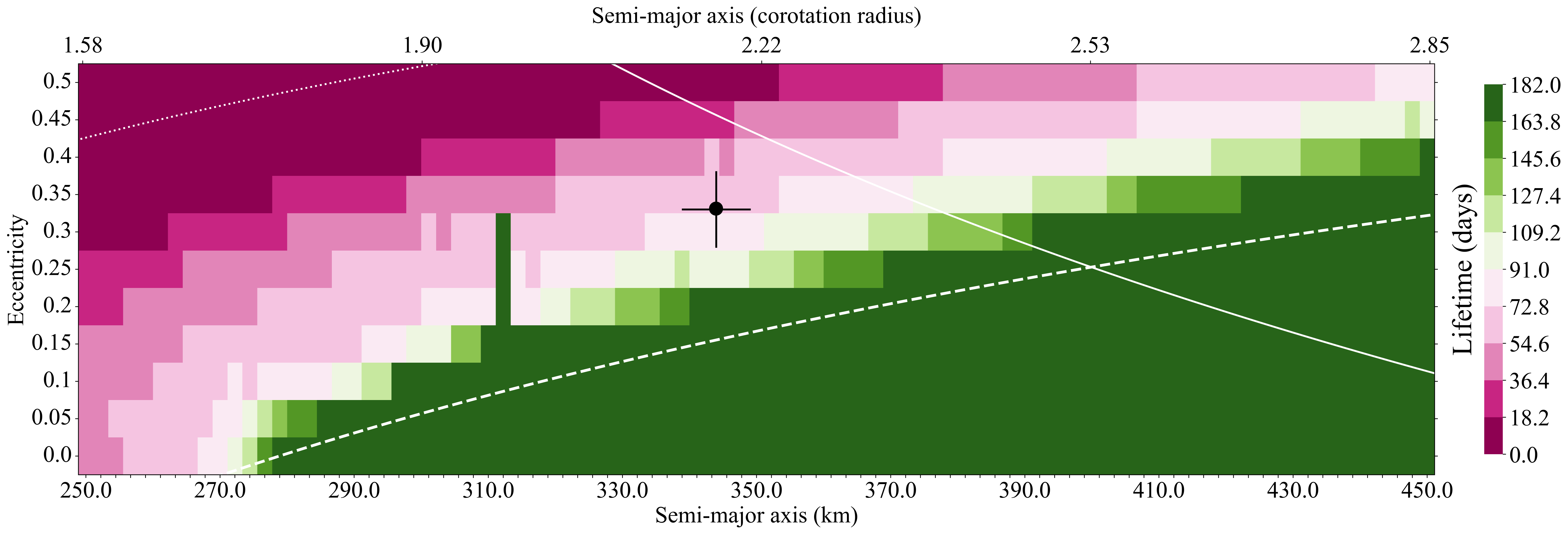}\label{afig:Mapatempo}
\end{center}
\caption{\label{fig:MapTempo} Map of the average lifetime in the region interior to the orbit of Gamma considering the full irregular shape of Alpha (using mascons). The color bar represents the lifetime of the particles in days.
All values higher than 182 days are indicated by the dark green color.
The black cross corresponds to the orbital elements of Delta with the respective error bars, while the black dot represents the nominal elements of Delta. The full white line indicates the collision line with Gamma. The dotted white line indicates the Roche limit pericenter curve. The dashed white line represents the chaotic limiting border given by the analytical expression of
\citet{lages2017chaotic, lages2018chaotic}.}
\end{figure*}

Figure \ref{fig:MapTempo} shows a map of the average lifetime
in the region interior to the orbit of Gamma.
We calculated the average lifetime by considering
all the particles of each box in our map.
We restricted the average lifetime up to 182 days. 
In the region below the limiting border, 
almost all particles have a
lifetime of at least 5 years, since this region includes
the surviving particles (see Fig. \ref{fig:MapaSemMassa}a).
The particles of the region around the black cross, which
corresponds to the orbital elements of Delta, have an average lifetime of a few months.
Thus, the region nearby the reported orbital elements of
Delta is unstable with a very short average lifetime expectation.

Therefore, we can conclude that the instability produced
by the full irregular shape of Alpha 
is mainly due to its elongated shape, represented by the ellipsoidal format, and also state 
that a model taking into account just the perturbation due to J$_2$ is not enough to represent 
the central body, Alpha. Thus, a new determination 
of Delta's orbital elements requires at least the perturbation
of Alpha as an ellipsoid to be considered.

\subsection{Massive body}
\label{massive}

For completeness, we verified how the stability of the system could be affected by taking into account the mass of Delta.
In this case, we considered the model of the full irregular shape of Alpha. 
We created a set of 1,000  initial
conditions to verify the interactions of a massive Delta with the system. 
Based on the ranges of the orbital elements of Delta, given in Table 1, we assumed a semi-major axis $a$
in the intervals of [334, 354] km, the eccentricity $e$ ranges from 0.23
to 0.43, and the inclinations ($I$) were randomly distributed in the 
interval of \ang{19} to \ang{57}).
The other angular orbital elements were randomly chosen
between \ang{0} and \ang{360}.
Each initial condition was simulated individually, 
and we evolved the system for up to five years.

\begin{figure*}
\begin{center}
\subfloat[]{\includegraphics*[trim = 0mm 0cm 0cm 0mm,
width=0.9\columnwidth]{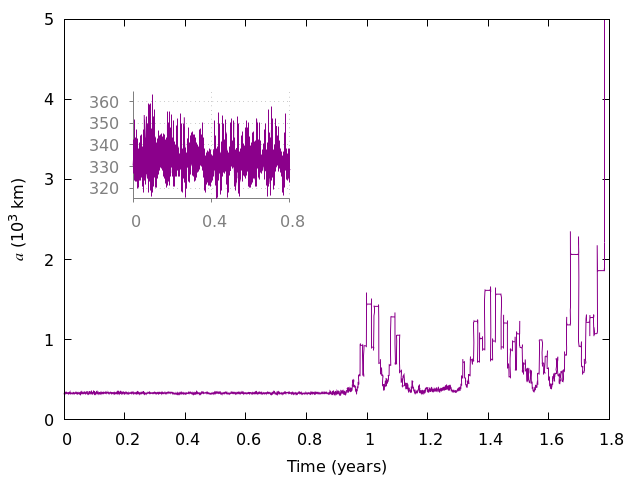}\label{afig:elema}} \hspace{1cm}
\subfloat[]{\includegraphics*[trim = 0mm 0cm 0cm 0mm,
width=0.9\columnwidth]{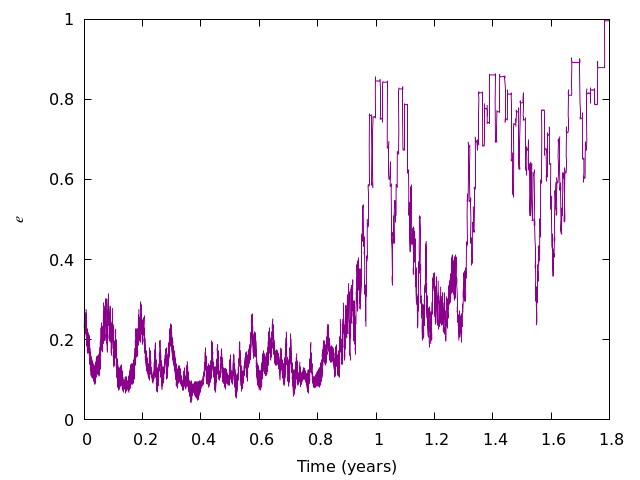}\label{afig:elemb}}\\ 
\subfloat[]{\includegraphics*[trim = 0mm 0cm 0cm 0mm,
width=0.9\columnwidth]{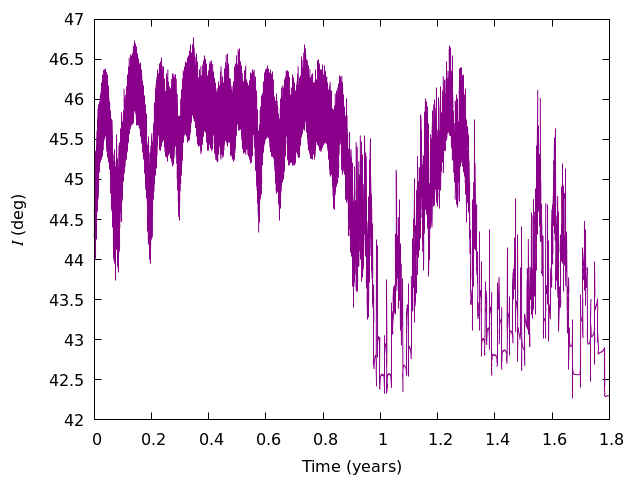}\label{afig:elemc}} \hspace{1cm}
\subfloat[]{\includegraphics*[trim = 0mm 0cm 0cm 0mm,
width=0.9\columnwidth]{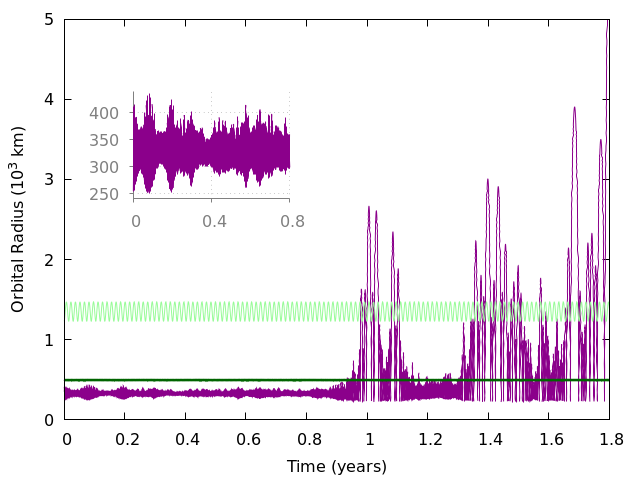}\label{afig:elemd}}
\end{center}
\caption{\label{fig:elem} Temporal evolution of the orbital elements of the massive Delta with the longest lifetime ($\sim 1.8$ years): (a) semi-major axis, (b) eccentricity, and (c) inclination. Panel (d) presents the orbital radius variations of Delta (magenta), Gamma (dark green), and Beta (green). The zooms in (a) and (d) show the evolution of the semi-major axis and orbital radius of Delta, respectively, for the first 300 days.}
\end{figure*}

None of the simulations remained stable.
From all the 1,000 initial conditions,  Delta survived
for at least 50 days in about 38\% of the cases, reaching 100 days in only 18\% of the cases. 
In just one particular simulation, 
Delta survived for about 1.8 years (Fig. \ref{fig:elem}). This longest survival case corresponds to
the initial orbital elements
$a$=$336$ km, and $e$=$0.23$, and $I$=$\ang{44.6}$. 
During the first year, Delta presents a variation
of about 50 km in the semi-major axis and \ang{3} in inclination, and 
after that, the variation increases (Fig. \ref{fig:elem}a,c). 
The eccentricity remains bounded to values smaller than 0.4 for the first year 
but shows a sharp increase after that (Fig. \ref{fig:elem}b). 
The orbital radius shows an erratic evolution (Fig. \ref{fig:elem}d). After
the first year, its orbit crosses the orbits
of Gamma and Beta, being finally ejected after 1.8 years.
This behaviour clearly shows a chaotic evolution of Delta. 

At the same time, the orbits of Gamma and Beta were not affected 
significantly by the gravitational
perturbation of Delta. 
That is  due to the fact that the masses of Gamma and Delta, and even Beta, 
are too small in comparison to the mass of the central body (Alpha), and the orbital inclination of Delta is very high in comparison to those of the other two bodies.
Thus their gravitational interaction
is  almost negligible.  These results suggest that Delta's unstable behavior is related to Alpha's irregular shape,  
and  we can assume that the mass of Delta 
does not affect the stability of the trajectories.

\section{Final comments}
\label{final}

Delta, the newly discovered satellite of (130) Elektra, 
was reported to have some uncertainties in its best orbital
fit. For this reason, we presented a stability analysis to
comprehend the environment where the moonlet 
is presumed to be \citep{refId0}. 

From the simulations and analyses presented in this work, we can conclude that:
\begin{itemize}
\item The region where Delta is supposed to be is fully unstable and is far from the nearest stable region;
\item The satellites Beta and Gamma do not play an important role in the dynamics of particles in highly inclined orbits that are interior to the orbit of Gamma;
\item The main source of the instability is the elongated shape of Alpha;
\item The determination of the orbital elements of Delta must take into account the gravitational potential of 
Alpha with  at least an ellipsoidal shape.
\end{itemize}

All the discussions of this work, whose objective was
to analyse the stability of the reported orbital elements of
the third new moonlet of (130) Elektra might contribute
to the understanding of the stability of the internal
region of the Gamma's orbit and the search for a new fit
for the orbital elements of Delta.


\section*{Acknowledgements}

This study was financed in part by the Brazilian Federal Agency for Support and Evaluation of Graduate Education (CAPES), in the scope of the Program CAPES-PrInt, process number 88887.310463/2018-00, International Cooperation Project number 3266, Fundação de Amparo à Pesquisa do Estado de São Paulo (FAPESP) - Proc. 2016/24561-0, Proc. 2019/23963-5 and Proc. 2022/01678-0, Conselho Nacional de Desenvolvimento Científico e Tecnológico (CNPq) - Proc. 305210/2018-1. RS acknowledges support by the DFG German Research Foundation (project 446102036).

\section*{ORCID iDs}
G. Valvano \orcidicon{0000-0002-7905-1788} \href{https://orcid.org/0000-0002-7905-1788}{https://orcid.org/0000-0002-7905-1788}\\
R. Machado Oliveira \orcidicon{0000-0002-6875-0508} \href{https://orcid.org/0000-0002-6875-0508}{https://orcid.org/0000-0002-6875-0508}\\
O. C. Winter \orcidicon{0000-0002-4901-3289} \href{https://orcid.org/0000-0002-4901-3289}{https://orcid.org/0000-0002-4901-3289}\\
R. Sfair \orcidicon{0000-0002-4939-013X} \href{https://orcid.org/0000-0002-4939-013X}{https://orcid.org/0000-0002-4939-013X}\\
G. Borderes-Motta \orcidicon{0000-0002-4680-8414} \href{https://orcid.org/0000-0002-4680-8414}{https://orcid.org/0000-0002-4680-8414}\\

\section*{Data Availability}
The data underlying this article will be shared upon reasonable request to the corresponding authors.



\bibliographystyle{mnras}
\bibliography{referencia} 

\begin{thebibliography}{}
\makeatletter
\relax
\def\mn@urlcharsother{\let\do\@makeother \do\$\do\&\do\#\do\^\do\_\do\%\do\~}
\def\mn@doi{\begingroup\mn@urlcharsother \@ifnextchar [ {\mn@doi@}
  {\mn@doi@[]}}
\def\mn@doi@[#1]#2{\def\@tempa{#1}\ifx\@tempa\@empty \href
  {http://dx.doi.org/#2} {doi:#2}\else \href {http://dx.doi.org/#2} {#1}\fi
  \endgroup}
\def\mn@eprint#1#2{\mn@eprint@#1:#2::\@nil}
\def\mn@eprint@arXiv#1{\href {http://arxiv.org/abs/#1} {{\tt arXiv:#1}}}
\def\mn@eprint@dblp#1{\href {http://dblp.uni-trier.de/rec/bibtex/#1.xml}
  {dblp:#1}}
\def\mn@eprint@#1:#2:#3:#4\@nil{\def\@tempa {#1}\def\@tempb {#2}\def\@tempc
  {#3}\ifx \@tempc \@empty \let \@tempc \@tempb \let \@tempb \@tempa \fi \ifx
  \@tempb \@empty \def\@tempb {arXiv}\fi \@ifundefined
  {mn@eprint@\@tempb}{\@tempb:\@tempc}{\expandafter \expandafter \csname
  mn@eprint@\@tempb\endcsname \expandafter{\@tempc}}}

\bibitem[\protect\citeauthoryear{Berdeu, Langlois  \& Vachier}{Berdeu
  et~al.}{2022}]{refId0}
Berdeu A.,  Langlois M.,   Vachier F.,  2022, \mn@doi [A\&A]
  {10.1051/0004-6361/202142623}, 658, L4

\bibitem[\protect\citeauthoryear{Borderes-Motta \& Winter}{Borderes-Motta \&
  Winter}{2017}]{borderes-motta18}
Borderes-Motta G.,  Winter O.~C.,  2017, \mn@doi [Monthly Notices of the Royal
  Astronomical Society] {10.1093/mnras/stx2958}, 474, 2452

\bibitem[\protect\citeauthoryear{{Geissler}, {Petit}, {Durda}, {Greenberg},
  {Bottke}, {Nolan}  \& {Moore}}{{Geissler} et~al.}{1996}]{Geissler1996}
{Geissler} P.,  {Petit} J.~M.,  {Durda} D.~D.,  {Greenberg} R.,  {Bottke} W.,
  {Nolan} M.,   {Moore} J.,  1996, \mn@doi [Icarus] {10.1006/icar.1996.0042},
  120, 140

\bibitem[\protect\citeauthoryear{Hanu{\v{s}}, Marchis, Viikinkoski, Yang  \&
  Kaasalainen}{Hanu{\v{s}} et~al.}{2017}]{hanuvs2017shape}
Hanu{\v{s}} J.,  Marchis F.,  Viikinkoski M.,  Yang B.,   Kaasalainen M.,
  2017, Astronomy \& Astrophysics, 599, A36

\bibitem[\protect\citeauthoryear{{Hu} \& {Scheeres}}{{Hu} \&
  {Scheeres}}{2004}]{huscheeres2004}
{Hu} W.,  {Scheeres} D.~J.,  2004, \mn@doi [PLANSS]
  {10.1016/j.pss.2004.01.003}, 52, 685

\bibitem[\protect\citeauthoryear{Lages, Shepelyansky  \& Shevchenko}{Lages
  et~al.}{2017}]{lages2017chaotic}
Lages J.,  Shepelyansky D.~L.,   Shevchenko I.~I.,  2017, The Astronomical
  Journal, 153, 272

\bibitem[\protect\citeauthoryear{Lages, Shevchenko  \& Rollin}{Lages
  et~al.}{2018}]{lages2018chaotic}
Lages J.,  Shevchenko I.~I.,   Rollin G.,  2018, Icarus, 307, 391

\bibitem[\protect\citeauthoryear{Marchis, Kaasalainen, Hom, Berthier, Enriquez,
  Hestroffer, Le~Mignant  \& De~Pater}{Marchis et~al.}{2006}]{marchis2006shape}
Marchis F.,  Kaasalainen M.,  Hom E.,  Berthier J.,  Enriquez J.,  Hestroffer
  D.,  Le~Mignant D.,   De~Pater I.,  2006, Icarus, 185, 39

\bibitem[\protect\citeauthoryear{Marchis, Descamps, Berthier, Hestroffer,
  Vachier, Baek, Harris  \& Nesvorn{\`y}}{Marchis
  et~al.}{2008}]{marchis2008main}
Marchis F.,  Descamps P.,  Berthier J.,  Hestroffer D.,  Vachier F.,  Baek M.,
  Harris A.~W.,   Nesvorn{\`y} D.,  2008, Icarus, 195, 295

\bibitem[\protect\citeauthoryear{Margot, Pravec, Taylor, Carry  \&
  Jacobson}{Margot et~al.}{2015}]{margot2015asteroid}
Margot J.-L.,  Pravec P.,  Taylor P.,  Carry B.,   Jacobson S.,  2015,
  Asteroids IV, 355, 374

\bibitem[\protect\citeauthoryear{{Merline}, {Tamblyn}, {Dumas}, {Close},
  {Chapman}  \& {Menard}}{{Merline} et~al.}{2003}]{2003IAUC.8183....1M}
{Merline} W.~J.,  {Tamblyn} P.~M.,  {Dumas} C.,  {Close} L.~M.,  {Chapman}
  C.~R.,   {Menard} F.,  2003, \iaucirc, \href
  {https://ui.adsabs.harvard.edu/abs/2003IAUC.8183....1M} {8183, 1}

\bibitem[\protect\citeauthoryear{Mirtich}{Mirtich}{1996}]{mir1996}
Mirtich B.,  1996, \mn@doi [Journal of Graphics Tools]
  {10.1080/10867651.1996.10487458}, 1, 31

\bibitem[\protect\citeauthoryear{Murray \& Dermott}{Murray \&
  Dermott}{1999}]{murray1999solar}
Murray C.~D.,  Dermott S.~F.,  1999, Solar system dynamics.
Cambridge university press

\bibitem[\protect\citeauthoryear{Vokrouhlick{\`y} \&
  Nesvorn{\`y}}{Vokrouhlick{\`y} \& Nesvorn{\`y}}{2008}]{vokrouhlicky2008pairs}
Vokrouhlick{\`y} D.,  Nesvorn{\`y} D.,  2008, The Astronomical Journal, 136,
  280

\bibitem[\protect\citeauthoryear{Winter, Valvano, Moura, Borderes-Motta,
  Amarante  \& Sfair}{Winter et~al.}{2020}]{winter2020asteroid}
Winter O.,  Valvano G.,  Moura T.,  Borderes-Motta G.,  Amarante A.,   Sfair
  R.,  2020, Monthly Notices of the Royal Astronomical Society, 492, 4437

\bibitem[\protect\citeauthoryear{Yang, Wahhaj, Beauvalet, Marchis, Dumas,
  Marsset, Nielsen  \& Vachier}{Yang et~al.}{2016}]{yang2016extreme}
Yang B.,  Wahhaj Z.,  Beauvalet L.,  Marchis F.,  Dumas C.,  Marsset M.,
  Nielsen E.,   Vachier F.,  2016, The Astrophysical Journal Letters, 820, L35

\makeatother
\end{thebibliography}








\bsp	
\label{lastpage}
\end{document}